\documentclass[10pt,conference]{IEEEtran} 
\IEEEoverridecommandlockouts
\usepackage{algorithmic}
\usepackage{graphicx}
\usepackage{textcomp}
\usepackage{xcolor}
\usepackage{balance}
\usepackage{url}
\usepackage{breakurl}

\usepackage{multirow}
\usepackage{enumitem}

\usepackage[numbers]{natbib}

\usepackage{hyperref}

\def\BibTeX{{\rm B\kern-.05em{\sc i\kern-.025em b}\kern-.08em
    T\kern-.1667em\lower.7ex\hbox{E}\kern-.125emX}}
\begin{document}

\renewcommand\citep{\cite}

\newcommand{\todo}[1]{}
\newcommand{\tomi}[1]{}

\newcommand{\note}[1]{}

\newcommand{\remove}[1]{}

\newenvironment{italicquote}
{\begin{quote}\itshape}
{\end{quote}}

%\renewenvironment{quote}
%  {\list{}{\rightmargin=0cm \leftmargin=\leftmargin}%
%   \item\relax}
%  {\endlist}

%\newcommand{\petri}[1]{{\color{Salmon1} PI: {#1}}}

\newcommand{\commentout}[1]{}

\newcommand{\univ}[1]{the University of Helsinki}
\newcommand{\Univ}[1]{The University of Helsinki}
%\renewcommand{\univ}[1]{\textit{Anonymised University}}
%the Hogwarts School of Witchcraft and Wizardry 
\newcommand{\UH}[1]{UH}

\title{Software startup within a university -- \\ producing industry-ready graduates\\
%{\footnotesize \textsuperscript{*}Note: Sub-titles are not captured in Xplore and
%should not be used}
%\thanks{Identify applicable funding agency here. If none, delete this.}
}

\makeatletter
\newcommand{\linebreakand}{%
  \end{@IEEEauthorhalign}
  \hfill\mbox{}\par
  \mbox{}\hfill\begin{@IEEEauthorhalign}
}
\makeatother

\author{\IEEEauthorblockN{Saara Tenhunen}
\IEEEauthorblockA{%\textit{dept. name of organization (of Aff.)} \\
\textit{\Univ{}}\\
Helsinki, Finland \\
saara.tenhunen@helsinki.fi}
\and
\IEEEauthorblockN{Tomi Männistö}
\IEEEauthorblockA{%\textit{dept. name of organization (of Aff.)} \\
\textit{\Univ{}}\\
Helsinki, Finland \\
tomi.mannisto@helsinki.fi}
\and
\IEEEauthorblockN{Petri Ihantola}
\IEEEauthorblockA{%\textit{dept. name of organization (of Aff.)} \\
\textit{\Univ{}}\\
Helsinki, Finland \\
petri.ihantola@helsinki.fi}
\linebreakand
\IEEEauthorblockN{Jami Kousa}
\IEEEauthorblockA{%\textit{dept. name of organization (of Aff.)} \\
\textit{\Univ{}}\\
Helsinki, Finland \\
jami.kousa@helsinki.fi}
\and
\IEEEauthorblockN{Matti Luukkainen}
\IEEEauthorblockA{%\textit{dept. name of organization (of Aff.)} \\
\textit{\Univ{}}\\
Helsinki, Finland \\
matti.luukkainen@helsinki.fi}
}

\maketitle

\begin{abstract}
Previous research has demonstrated that preparing students for life in software engineering is not a trivial task. Authentic learning experiences are challenging to provide, and there are gaps between what students have done at the university and what they are expected to master when getting into the industry after graduation. To address this challenge, we present a novel way of teaching industry-relevant skills in a university-led internal software startup called Software Development Academy (SDA). In addition to describing the SDA concept in detail, we have investigated what educational aspects characterise SDA and how it compares to capstone projects. The questions are answered based on  15 semi-structured interviews with alumni of SDA. Working with production-quality software and having a wide range of responsibilities were perceived as the most integral aspects of SDA and provided students with a comprehensive skill set for the future. 

\end{abstract}


\section{Introduction}
\label{sec:intro}

Universities are responsible for equipping new graduates with sufficient skills and abilities to enter working life. In software engineering, this includes providing the basic theoretical understanding in computer science \citep{ieeeCS, ieeeSE}, technical competencies and knowledge demanded by the industry \citep{radermacher2014investigating, garousi2019closing}, as well as soft skills~\citep{ahmed2012evaluating}. 

\subsubsection*{Soft skills}

The need for soft skills and work-life skills, in general, is ever more important in software engineering because of the rapid technical development of the field. Employers require universities to produce graduates with appropriate ``employability skills`` such as communication and self-management \citep{johns2013simulating}. The emphasis of each programme is unique, but some of the commonly mentioned soft skills that students should possess after completing their studies are teamwork \citep{keogh2007scalable, ziv2010capstone, delgado2017evolving, marques2017enhancing, iacob2019exploring}, client negotiation skills \citep{keogh2007scalable}, project management \citep{haddad2013one} and general ability to function in the software engineering industry \citep{ziv2010capstone, mahnic2011capstone}. These skills are particularly central in project courses, often provided at the end of a study program and where students can demonstrate their cumulative knowledge (i.e., capstone courses) \citep{ieeeCS, ieeeSE}.

In a systematic literature review (SLR) on the skill gaps of computer science and software engineering graduates, written communication tied with oral communication as the most commonly identified knowledge deficiency \citep{radermacher2013gaps}, with project management coming in third. Another SLR on the gap between SE education and industry expectations found that professional practice, containing professionalism, group dynamics and communication skills, is perceived as highly important in the industry \citep{garousi2019closing}. Yet it still presents a high knowledge deficiency and therefore is something educators should pay close attention to \citep{garousi2019closing}. This is also emphasised in the latest Guide to Software Engineering Body of Knowledge (SWEBOK) \citet{swebok3}, where a new knowledge area called Software Engineering Professional Practice was introduced and further divided into sub-areas called professionalism, group dynamics and psychology, and communication skills. 

\subsubsection*{Technical skills} 

Regarding technical skills, the knowledge deficiencies may not be as imminent or all-encompassing as with soft skills \citep{radermacher2013gaps,begel2008novice, stevens2016industry}. Whereas technical foundations are often solid, many industry representatives acknowledge the need to extend the specific technical skills \citep{stevens2016industry}. The technical skills that industry representatives report the graduates missing most often relate to software development tools and configuration management \citep{begel2008novice, radermacher2013gaps, radermacher2014investigating, garousi2019closing}. Hired graduates likely have not used software tools in a production environment before and might lack an inherent understanding of why a production environment should be used \citep{radermacher2014investigating}. %Configuration management has been identified as being of high importance in the industry while being the topic with the highest gap between industry expectation and software engineering education \citep{garousi2019closing}.

\subsubsection*{Capstone project courses} While technical foundations are covered in lower-level courses, many institutions hold team-based capstone project courses to teach soft skills and to ensure students are ready to apply technical knowledge in real life \citep{ziv2010capstone, majanoja2018reflections, panicker2020exposing}. In these large projects, students are expected to implement software systems and experience stages of the software development life-cycle from requirements solicitation to software maintenance \citep{buffardi}. Moreover, capstone projects are often seen as invaluable in teaching soft skills. This includes teamwork \citep{keogh2007scalable, venson2016academy}, verbal and written communication \citep{watkins2010competitive}, time management \citep{dupuis2010experiments}, problem solving \citep{majanoja2018reflections} and project management \citep{haddad2013one}. For students, a capstone project typically represents a culmination of their studies and is one of the last milestones before graduation \citep{ieeeCS}.

Projects in software engineering programs typically last one or two semesters~\citep{ikonen2009discovering, adams2016collaboration, schneider2020adopting} and are sometimes organised with industry clients \citep{fornaro2007reflections, isomottonen2008value, majanoja2018reflections, spichkova2019industry}. A literature review by Dugan~\cite{dugan2011survey} divided capstone courses into industry projects, community service, gamified course environments, other simulated course experiences, and research projects.

\subsection{Research problem} 

Although capstone courses help students to gain relevant skills and provide an important sneak peek into the work-life in software engineering, the essence of what work in a professional software team is all about is hard to capture. Moreover, previous research has demonstrated that preparing students for life in software engineering is not a trivial task. There are still clear gaps between what students have done in their software engineering projects at the university level and what they are expected to do in the industry \citep{begel2008novice, radermacher2013gaps, radermacher2014investigating, garousi2019closing}. %Thus, new ways of teaching and learning authentic software engineering are interesting to explore.

Several studies (see \citep{tenhunen2022capstones}) have pointed out that a typical capstone course usually progresses from an idea to a working prototype and often does not include deployment, operation and maintenance of the software in a production environment. This is a big challenge since running and maintaining an application in production requires its own set of techniques and skills, and in real life, most software development is done in the context of operational systems.

Since 2017, \univ{} has been running an internal, non-profit software startup called SDA\footnote{``\emph{An internal software startup operates within the corporation and takes responsibility for everything from finding a business idea to developing a new product and introducing it to market. [...] Compared to the traditional R\&D activities of larger companies, an internal software startup develops products or services faster and with higher market orientation}.''~\cite{unterkalmsteiner2016software}}. SDA is comprised of computer science students and is led by faculty. Selected students work in SDA for one year, consisting of part-time work during the academic term and full-time work during the summer. During the year, each student has a wide range of software development, operation and maintenance responsibilities. Although SDA was not initially targeted for teaching purposes, it has turned out to be an interesting educational tool.

In this study, we describe the concept of SDA in detail. We interviewed SDA alumni to find out how they describe their learning experiences in SDA and how they compare SDA to other learning experiences. More specifically, we will answer the following research questions: 
\begin{itemize}
    \item[] 
     {\hspace{-10pt}\textbf{RQ1.} How  is SDA characterised as a learning experience? 
    \item[]  %\emph
    {\hspace{-10pt}\textbf{RQ2.} How do traditional capstone courses differ from SDA?}
    \item[]  %\emph
    {\hspace{-10pt}\textbf{RQ3.} What are the deficiencies of SDA? }}
\end{itemize}

RQ1 is motivated by how understanding the educational aspects behind internal software startups helps other educators introduce similar constructs and identify what learning opportunities might be interesting to add, e.g., in capstone courses. RQ2 complements this by explicitly comparing SDA and capstone courses. Finally, RQ3 focus on the potential downsides of our approach.

The rest of the paper is structured as follows. The context, i.e., the SDA concept, is described in Section~\ref{sec:case_introduction}. The details of the conducted case study are provided in Section~\ref{sec:methods}. The comparison results of SDA and project courses are provided in Section~\ref{sec:results}. The relevance and validity of the identified differences and student experiences are discussed, and conclusions are drawn in Section~\ref{sec:discussion}.

\section{Software Development Academy}
\label{sec:case_introduction}

\begin{table*}[!ht]
\renewcommand{\arraystretch}{1.25}
\caption{Main aspects of capstone projects and a university-lead internal software startup (SDA)\label{tab:comparison_capstone_sda}}
\begin{tabular}{p{0.22\linewidth} p{0.45\linewidth} p{0.28\linewidth}}

\textbf{Feature}~\citep{tenhunen2022capstones} & \textbf{A typical capstone course}~\citep{tenhunen2022capstones} & \textbf{Internal software startup (SDA)} \\
&(a software engineering capstone in \univ{})
%the University of Helsinki) 
& \\
\hline
Duration & Few months to one year (14 weeks) & 1--2 years \\
Intensity & Few hours to 40 hours (15 hours per week) & 8 months half-time, 4 months full-time  \\
Salary/Employee status & No (No) & Yes \\
%Employee status & No & Yes \\
Clients & Little over half of the courses have external clients from the industry or from other units in the university  (external clients) & External clients from other units in the university\\
Project sourcing~& Industry-proposed, teacher-generated or student-proposed (Proposed~by~the~clients) & Manager finds suitable projects that are based on a real need \\
Software phases & Mostly greenfield coding: from idea to a robust proof-of-concept (Mostly greenfield, some existing project, rarely in production use)  & Existing projects in production use \\
Team size & 1--35 students per team, generally 5--6 students (5-7~students) & 4--10 students and one manager \\
Team composition & One team of students working on one project, generally no interaction with other teams (One team, no interaction with other teams) & Several, partially overlapping subteams for different projects \\
Continuous feedback and guidance & Available mostly from faculty, sometimes from industry or more experienced students. (From faculty about the process and the client about the product) & Available from faculty and more experienced students \\ 
Student selection & Generally students with certain pre-requisite courses completed or of a certain class in their studies (Students with pre-requisites) & Selective recruitment process \\
Technologies used & Varies a lot, some have a common stack, some do not (Stack decided mostly by the students, sometimes decided by the client) & Common stack across projects \\
\hline
\end{tabular}
\end{table*}

Software Development Academy (SDA) is an internal, non-profit software startup comprised of computer science students led by faculty within \univ{} (\UH{}). SDA develops and maintains several educational administrative applications for the use of the entire university. SDA was founded in 2017 and has undergone various changes over the last five years to suit the needs of the team and its stakeholders. The framework here represents SDA as it was in spring 2022 and has been generalised as much as possible.

\subsection{Staff}

The team size of SDA %is not fixed and 
varies between 4--10 students depending on the need. By default, each student is given a one-year contract. Rapid circulation of staff ensures that no software is left as a responsibility of a single employee for too long. Knowing that the time at SDA is limited forces the students to demonstrate their work more often and pass any relevant information to the remaining team. It also aims to motivate the students to continue their studies to gain further employment after their time at SDA. Student staff members are all given the same salary at the corresponding experience level.

No specific year or phase of studies qualifies a student to be hired. Instead, the competencies of the student are at the centre. Few have had pre-existing degrees and several years of work experience behind them, whereas SDA is their first real job for some. Some have only recently started their studies, and some are finishing their Master's degrees. Through their wide responsibilities as a teacher in the Computer Science Department, the manager of SDA has a very good overview of the entire student population in the CS programs. They are responsible for several compulsory and elective courses on software engineering and related technologies. This overview gives the manager a general idea of who might be a potential recruit for SDA. In addition, the students working in SDA can propose or second any selection based on their experiences with their fellow students. The recruitment process, therefore, has remained a fairly straightforward one. Generally, one interview with a potential recruit has been sufficient to determine whether that person gets hired.

All students work at SDA half-time during the academic year and are expected to continue their studies while working. SDA differs from traditional summer internships in the sense that students are hired throughout the year, and normally only one new team member starts at a time. Often, when one student member leaves, a replacement for them is hired. A sufficient overlapping period is also reserved, where the leaving staff member hands over the responsibilities  and software they have been working on to the new staff member. This procedure also ensures that the existing staff members are not overwhelmed by the need to guide and mentor many new team members at once during summertime, which is a risk with hiring plenty of summer interns. The spread of starting dates also balances out the proportion of more and less experienced developers in the team at any given moment.

SDA is managed by a single faculty member who simultaneously is a lecturer at \univ{}. Approximately 30\% of their monthly work time is allocated to SDA-related issues and paid by the faculty. Their main responsibilities are supervising the team of students, participating in client meetings, negotiating for funding (even the internal clients are expected to pay for the software) and starting new projects, keeping updated on the status of each project and recruiting new students. The manager is very involved in the daily life of SDA. In addition to the manager, SDA has one faculty member, working partly as a software developer and providing technical continuity for the team. But their main duty is teaching, which reduces the amount of time allocated to helping the students in SDA.
\note{Siirsin ylläolevan pätkän tähän tuolta tämän subsectionin lopusta. Ja muokkasin pikkusen seuraavan kappaleen ekaa lausetta. Oli mun mielestä parempi flow näin, mutta saa toki muuttaa takas, jos huononi.} 

Leading SDA poses various challenges for the managing faculty member, compared to working as a lecturer. Firstly, it is essential that the manager keeps up with the recent technological trends, which is not necessarily easy for someone working within an academic institution (see, e.g.~\cite{luukkainen2012}). It is easy for the most talented students to get employed in the industry, so recruiting suitable candidates requires a position where potential candidates can be proactively scouted and hired well ahead of time. The manager also needs to be very careful to pick only those projects that fit SDA well. The technology, the domain, the user base and the project ownership need to be well aligned with the SDA skill set and goals. The manager also needs to market and promote SDA within the university to secure its position and funding in the future. 

\subsection{Software Projects and Clients}

Most of the projects that SDA has undertaken are web applications uniquely specific to \univ{}. The user base for the software varies from a few dozen faculty members to the entire student and teacher body of the university. The largest software, in terms of its user base, is a course feedback system used by every teacher and student at \univ{}, having more than 1000 daily users on the busiest days. Other software includes a self-assessment form provided for the study programmes in the university, an analytics tool forcomparing student cohorts, study programmes and courses, and a tool automising course completion registrations.

At the moment of writing this, there are four client organisations with applications under development and maintenance. All four are organisations within \univ{}, and SDA manages several software applications for two of these.
The number of applications has grown over the years as earlier launches have been successful, and at the moment, it totals 10. Out of these, a little over half have been in active development within the past year, meaning that they have had large new features implemented. The remaining four are smaller software in fairly low-key maintenance mode and rarely require attention from the staff. All the software developed in SDA is open source\footnote{Most of the projects developed and maintained by SDA can be found in \url{https://github.com/UniversityOfHelsinkiCS}}. Having the source code open to the public also allows the student members to use it as a reference point when applying for jobs later. %after their time at SDA.

The manager of SDA needs a dose of business sense to pick the right projects for development. Besides a technological fit, a project must have a clear business owner with a strong connection to the user base and authority for decision-making. It is extremely important that there is not much bureaucracy around the project that would slow down the progress. 

\subsection{Organization of work}

SDA is divided into several subteams of approximately 2--5 students. Each subteam is responsible for one software that is in the active development phase. There are no working times set in stone for any of the subteams, and they are very self-organized when it comes to their ways of working. Since the working hours are not fixed, the students are able to make their own schedules so that they can attend any lectures, group tasks or practice sessions they might have in their studies. The entire staff shares one weekly meeting at the end of the week to summarise the past week and share ideas and thoughts.

Each subteam usually has at least one senior member who has already worked with the product for some months. Besides the development, a senior member guides the other team members, ensuring that information gets passed from one student generation to the next. The junior members of a subteam are gradually taking wider responsibilities, becoming the new seniors after some months of working.

When the interviews were conducted, the team worked locally in the office on campus and remotely from home. Each student was free to choose where and when they wish to work. Instant messaging platforms are in very active use at SDA for both off- and on-topic conversations. It is worth mentioning that the global COVID-19 pandemic significantly affected SDA's work practices. Between spring 2020 and fall 2021, the entire team was forced to work remotely, whereas before the pandemic, students worked mostly on campus.

SDA has no separate account managers to conduct communication with the clients. While the manager maintains the big picture of the developed software, each staff member is also responsible for communicating directly with the client about the software they are developing. This includes the whole process of finding out the requirements of the client and creating and presenting a solution. Each project that is in active development has weekly meetings where the Product Owner from the client side, the developers of the software and the SDA manager are present. The meetings entail going through future tasks and lately implemented fixes and features.

Despite having one week cycle in the customer meetings, the subteams are not following fixed iterations or sprints in the development, which are typical in Agile development processes such as in Scrum \citep{SchBee02}. Instead, the organization of the development resembles a Lean process \citep{PoppendieckPoppendieck03}, where the user-level features are implemented and taken to production one by one, as quickly as possible, trying to minimize the lead time \citep{liker2003toyota} that is, the time from idea to production.

\subsection{Technologies}

The technology stack used at SDA has been consolidated over the years. React is used in the frontend and Node.js in the backend. The logic for these choices goes back to the initiation of SDA in 2017. The combination of React and Node.js was seemingly becoming the ``go-to stack`` in the web development industry, which made it a natural selection to be the building blocks of SDA software.

All software is containerised and runs on virtual machines located on servers, which are maintained by the university. Each software uses the same version control system, has similarly configured CI/CD pipelines and employs end-to-end testing. The consolidated stack helps the team in many ways. Firstly, staff members can fairly easily fluctuate between different projects or get a grasp of multiple software at once. Different subteams working on different software can easily share best practices with each other. It also allows the staff to gain knowledge of one commonly used stack fairly deeply rather than scratching the surface of multiple technologies. At this level of experience, the continuity in the technologies used from project to project is integral. The skills of junior developers are not wide enough for them to jump from one project to another within one year if each project is done with different technologies.

\subsection{SDA and Software Engineering Capstone Course}

\note{Lisäsin tähän "Capstone" kun se on muuallakin tässä artikkelissa lisätty. Oikea kurssin nimi taitaa olla ilman sitä, mutta ehkä selkeämpi olla yhtenäinen jompaankumpaan suuntaan. Tietenkin kaikki voi findilla ettiä ja korvata olemaan ilman Capstonea myös.  -Saara\\Olin jättänyt pois, koska kurssin nimessä ei ole sanaa capstone ja ison kirjaintan otsikkomainen käyttö antaa ymmärtää, että kyseessä ni "jonkin" nimi. Capstonen lisääminen kuitenkin auttaisi lukijaa -Petri}
In addition to SDA, \univ{} also provides a mandatory capstone project course, called Software Engineering Capstone Project, for the students in Computer Science Bachelor Program. Students are expected to take the course after passing most of the Bachelor-level courses, typically in their third year. The course lasts for 14 weeks, and students are assigned into teams of 5--7 students based on their self-assessed skills and preferences regarding project topics. The clients for the projects comprise local businesses as well as research groups and various departments of %the 
\univ{}. All clients are external, and none of the teaching staff of the course act as clients. The manager and founder of the Software Development Academy, the fifth author in this research, has also been the responsible teacher for the course since 2009. Comparison of the course designs based on the taxonomy of the features of capstone projects \citep{tenhunen2022capstones} is provided in Table~\ref{tab:comparison_capstone_sda}.

\section{Methods}
\label{sec:methods}

\subsection{Participants}

In March 2022, there was a total of 17 SDA alumni. They were all contacted and asked if they would like to participate in an interview about SDA. There were 15 willing to do so, from whom 14 were employed full-time in the software industry when the interviews were conducted.\footnote{One interviewee was not employed at the time of the interview due to their own choice. At the time of writing this article, they too were employed in the software engineering industry.} The rest of the participant demographics are summarised in Table~\ref{tab:background-students}. In order to keep interviewees' individual answers anonymous  
quotations taken from the interviews are not relatable to any specific interviewee or their demographics.

\begin{table}[!th]
\caption{Demographics\label{tab:background-students}}
\begin{tabular}{p{3.5cm} l r}
\textbf{Variable} & \textbf{value} & \textbf{n} \\ \hline
\multirow{3}{=}{Professional software engineering experience before SDA} &None & 6 \\ 
&$<$ 0.5 years & 4 \\
&0.5 year - 2 years & 3 \\
&$>$ 2 years & 2 \\ \hline
\multirow{3}{=}{Phase of studies when starting at SDA} &2nd year Bachelor student & 5  \\
&3rd year Bachelor student & 4 \\
&1st year Master's student & 4 \\
&2nd year Master's student & 1 \\ \hline
\multirow{2}{=}{Software engineering capstone project done before SDA} &Yes & 9  \\
&No & 6 \\ \hline
\multirow{3}{=}{Worked partly or entirely remotely at SDA, due to the COVID-19 pandemic} &Yes & 5  \\
&No & 10 \\ \\ \hline
%\multirow{2}{=}{Employed in software industry after SDA} \\ &Yes & 14  \\
%&No* & 1 \\
%\\

\end{tabular}
\end{table}

\subsection{Positionality}
\label{sec:positionality}

Especially in qualitative research, authors' identities and perspectives on the research may affect the research process and outcomes~\cite{secules2021positionality}. Thus, it is essential to be transparent about our position. SDA was established and is currently led by Dr Luukkainen. In addition, he has taught the Software Engineering Capstone Course at \univ{} since 2009. Interviews were conducted by Ms Tenhunen (n=8) and Mr Kousa (n=7), who have also worked in SDA. The preliminary findings of this work were reported in the Master's thesis by Ms Tenhunen.

\subsection{Interviews}

Most of the interviews \todo{(n=XXX)} were conducted online via Zoom with video on, while only few\todo{(n=XXX)} were conducted face to face. All interviews started with questions related to the background of the interviewees before their time at SDA. The rest of the interviews were organised under four main themes (see the Appendix for further details):

\begin{enumerate}
\item Experiences from Software Development Academy 
\item Comparing SDA with project courses 
\item Learning useful skills 
\item Work-life experience 
\end{enumerate}

 The idea was to gain an idea of the level of their skills and understanding of the industry prior to working at SDA. The first main theme was their experiences regarding SDA. The questions covered the responsibilities that the students had at SDA as well as their perception of SDA. The second theme the interviewees were asked to compare SDA to project courses they have taken. These questions were asked in order to find out not only if there exist any differences between regular project courses and SDA but also whether SDA could be replaced by or expanded into a course. The third theme was specifically about learning the skills relevant to work-life, including questions on the industry-relevant skills the interviewees had learned at SDA and elsewhere at the university. The fourth theme focused on the time after working at SDA, e.g., by asking about the potential differences between SDA and the current place of employment. 

The interviews were conducted in Finnish (i.e., the native language of the interviewees) and lasted between 30 minutes and 2 hours. All the interviews were recorded and transcribed. \note{Poistin seuraavan ja lisäsin vastaavan inforn positionality kappaleeseen, jonka siirsin tänne menetelmiin: "The first author of this research conducted eight interviews, and the fourth author did the remaining seven."}

\subsection{Analysis}

The interview questions were wide-scoped, and the research topics were approached multiple times. Thematic analysis was chosen as a commonly used method for describing, analysing and reporting themes and patterns in data \citep{alhojailan2012thematic}. All the interviews were analysed when seeking excerpts related to the educational aspects (RQ1) and downsides of the approach (RQ3). The answer to RQ2 was based on analysing the answers in themes three and four (See the Appendix).

The analysis was conducted by the first author. The essential excerpts of each answer were identified and coded on the datasheet, and the main topics were obtained by grouping the topics iteratively. The final themes and related excerpts were discussed with the other authors, but interrater reliabilities were not calculated.

\section{Results}
\label{sec:results}

\subsection{Educational aspects (RQ1)}\label{educational_aspects}
\label{sec:aspects}

The educational aspects we identified in the analysis of the interview data are listed in Table~\ref{tab:sda_dimensions} and will be explained in detail in the following.

\begin{table}[b]
%\scriptsize
\caption{Identified educational aspects (topics) and how many interviewees raised the topic (n).
%and challenges in SDA
\label{tab:sda_dimensions}}
\begin{tabular}{p{0.7\linewidth}
p{0.2\linewidth}
}
%\toprule
%\hline
\textbf{Educational aspects} & \textbf{n} \\ 
%\midrule 
\hline
%Educational aspect \\
%\midrule 
%\hline
Working with software in production & 15 \\
Wide responsibilities and autonomy  & 15 \\
Employee status and salary  & 14 \\
Strong community and networking possibilities  & 13 \\
Easy integration with studies & 11 \\
Selectiveness of team members & 11 \\
One-year duration & 10 \\
Working with external stakeholders & 6\\
%\midrule 
%Perceived challenges \\
%\midrule 
%Lack of clear seniority & 9 \\
%Asynchronized working schedules & 6 \\
%\bottomrule
\hline
\end{tabular}
\end{table}

\note{Muutin tän tähän muotoon. Se pointti tosiaan oli enemmän se että Toskan softa on tuotannossa eikä leluprojekti}
\subsubsection{Working with software in production}

The most commonly mentioned positive aspect of SDA was that the software goes to production and actual use by the customers. All 15 interviewees made a note of this, especially compared to software projects they had worked on during their studies. Project-based courses commonly include developing software from an idea to a prototype but not working in a production environment (see Section~\ref{sec:intro}). 

The fact that the developed applications had real clients was related to multiple subcategories of comments. Firstly, working in production forced students to focus on the quality of their work. Even if the software did not have millions of users\todo{(add some)}, it was never-the-less expected to run without interruptions. Any changes introduced to the software should not leave it broken. Secondly, students learned how to handle and maintain CI/CD pipelines and infrastructure resources, keeping in mind that the software is in production use at all times. Thirdly, several interviewees indicated that delivering clear value for the end-users increased their motivation.  Finally, students could say in a job interview that they had worked on production-level software for a year.

\begin{italicquote}
``(Developed features) to production as fast as possible. That is something that you don't necessarily get to experience in other environments. And you don't understand the value of that until you witness it first-hand. The long-term nature of the development is important. It is not enough that the application works for six weeks and then explodes on the eighth (week). It adds seriousness to the process.``
\end{italicquote}
\vspace{.4mm}
\begin{italicquote}
``The sense of meaning in work and releases in SDA is high. There are actual users who are happy with the updates made in the software.``
\end{italicquote}

\subsubsection{Wide responsibilities and autonomy}

All the interviewees considered it important that the students were given wide responsibilities in contrast to narrow and simple tasks. This made students feel that they understood software development better as a whole. Most interviewees described how they started at SDA by working on simple front-end issues, such as minor bugs or feature requests. From there on, they progressed deeper into the backend of the software and built larger and more comprehensive features. Building new features often included refactoring older parts of the codebase and making changes to existing databases. Moreover, most interviewees ended up taking on responsibilities related to the infrastructure that the software runs on. This meant working on the deployment pipelines or setting up software on new virtual machines. Some spent much time integrating the developed software into other systems used at the university. By the end of their year, an average student had a lead role in at least one software and acted as a mentor to newcomers. The wide variety of tasks and areas enabled students to gain precious skills for their future work life.

\begin{italicquote}
``You get a really vast experience from different areas of software development fast. You get to do everything. And you get to do it almost immediately.``
\end{italicquote}
\vspace{.4mm}
\begin{italicquote}
``It is not that I didn't learn any new skills at SDA. It is that there are so many skills, I think it is worthless to start listing all of them.``
\end{italicquote}
\vspace{.4mm}
\begin{italicquote}
``Coding. Infrastructure. Designing software architectures. Testing. People skills. Working with colleagues. Trunk-based development. That's about it. All sorts of coding. Infrastructure, databases, message systems, end-to-end testing. A vast set of skills and knowledge.``
\end{italicquote}

Interviewees (who currently work in the industry) mentioned that taking up such a high responsibility and a variety of tasks in one year was a rare opportunity, especially when compared with larger companies, teams or software, where the responsibilities tend to be much more refined and focused. Some interviewees also explained that this kind of variety of tasks created the possibility to test various things in a working context and only then decide which direction to specialise in the future.

\begin{italicquote}
``I could not perform in my current job as a software developer if I hadn't been to SDA first. I would not survive my tasks. Entering working life right after the Software Engineering Capstone Project would have made me cry.``
\end{italicquote}

Along with wide responsibilities, a few interviewees specifically mentioned high autonomy regarding software development choices and practices as positive. Being able, and also having the responsibility, to choose how to implement a feature was satisfying and an important motivational factor. However, it is worth mentioning that the high level of autonomy was perceived not only as positive. Some interviewees noted that while having freedom in their work was nice, autonomy combined with relatively inexperienced developers led to poor choices or less optimal ways of implementing a feature, tool or configuration. 

\begin{italicquote}
``Even though there are some guidelines and rules on how people do things in SDA, there are so so so much fewer of those things than in some companies. That might show in the way people sometimes make insane choices when it comes to development. For instance, someone might code something that has no use. Or focus on something funny. This, of course, is nice and fun, but might not give so much to the product being developed.`` 
\end{italicquote}

\subsubsection{Employee status and salary}

One of the significant differences between project courses and SDA is the salary. Even though the salary is not high, the interviewees reported it as one of the primary motivators to join SDA. With salary also comes the job contract, which ties the student members to the team on a different level than traditional capstone projects. 

\begin{italicquote}
``It is a good aspect of SDA that you work there and get money. I think the outcome would be very different if SDA were a course worth 25 credits without money.``
\end{italicquote}

For some, being an employee of the university was valuable as such. They reflected on how it was interesting to be in the ``inner circle`` at the university. They got the official status of saying they were working as developers and had a dedicated working space at the university. Most mention that they have found the extrinsic value of the employee status when applying for employment after their time in SDA. A year's worth of industry-relevant work experience using modern technologies gave them a head-start in their job search.

\subsubsection{Strong community and networking possibilities}

Almost all interviewees brought up the community, which has formed at SDA over the years, as a positive factor. Some explained how they had created real friendships with their colleagues, and others how they had enjoyed having like-minded people to work with. This not only made their time at SDA more enjoyable but also affected their future employment. In the interviews, 3 out of 15 mentioned finding employment directly due to the relationships they built at SDA. Few others had been in job interviews arranged by people at SDA, even though they did not end up in those positions. One interviewee mentioned that it was nice to know people working in the same industry but not in the same company as they currently are. This allowed them to gain perspective on how others do things in the industry.

\begin{italicquote}
``Sense of community and team spirit is unusually strong at SDA. It has also shown after I left SDA. SDA has a strong community, and people care about each other and take care of each other.``
\end{italicquote}

\subsubsection{Easy integration with studies}

One recurring topic in the interviews was the convenience and flexibility of working at SDA. The location of the office at the campus, right alongside auditoriums and classrooms where the CS courses are held, was perceived as very handy. Some interviewees mentioned working a few hours between lectures as possible. Not having to work strictly from nine to five was convenient for many. The general atmosphere also encouraged them to study, and few found their Master's thesis subjects within SDA.

\begin{italicquote}
``SDA fits into the studies well. Studies and SDA are really easy to integrate into one another. While I was doing SDA things, I was also doing school things. For instance, project work could be of both. In that sense, it was an efficient use of my time. I also gained a subject for my thesis from SDA and peer support for doing it. That is something that was not available elsewhere.``
\end{italicquote}

Despite the convenient arrangements,  part-time work took time from other activities. When asked about the study progress during SDA, 5 out of the 15 said that working at SDA was preoccupied and slowed down their studies. Two interviewees mentioned that it was hard to estimate the effect since they were other factors, such as the COVID-19 pandemic. Six felt that they progressed as planned. Some of these six were finishing their studies and felt that SDA did not have any effect in one way or another. Two mentioned that their time at SDA boosted their pace in studies as the environment was encouraging.

The ones who progressed as planned, or better, mentioned that flexibility with the working hours, the general attitude towards the importance of studies in the team, and the close relation to the university community helped them to maintain their pace of progress.

\subsubsection{Selectiveness of team members}

When asked about the negative aspects of SDA, few raised exclusivity as an issue. Some felt that it was an unfair advantage that few selected members of the student body were given, and not everyone was able to have such an experience during their studies. Others, however, raised exclusivity up but in a positive light. Since the students were specifically hired to SDA, there was a good chance that they were motivated and fit into the environment well. This not only made the team more productive but also the working experience more enjoyable. With capstone projects, many interviewees felt that since the course was compulsory, the groups were often random collections of students, negatively affecting the team's overall performance. %The selective recruitment process also comes with a risk. 
As with any company hiring, recruitment to SDA contains the risk of the wrong person getting hired. Some interviewees perceived this as a challenge at SDA.

\subsubsection{One-year duration}

Capstones and other project-based courses tend to last only one semester, which does not necessarily give students sufficient experience in software engineering. Many interviewees shared this opinion regarding the Software Engineering Capstone Project at \univ{}. SDA, however, lasts for a year, including full-time work in the summer. This provided ample ground for students to build their skills.

\begin{italicquote}
``In SDA, you have a lot more time, and it gives you the possibility to dive deeper into the projects. Even though the Software Engineering Capstone Project is 200 hours, it still is a scratch on the surface when it comes to software development. After all, you are supposed to fit the whole software lifecycle in that 200 hours.``
\end{italicquote}

Nevertheless, some criticised the one-year duration as too short of a time to experience everything they wanted.

\begin{italicquote}
``I, of course, understand why students get to spend only one year in SDA. But still, I feel that the first year in a new job is usually some sort of an orientation period. ... That is how long it takes to make yourself comfortable and acquainted with any team, no matter how good the team is and how well you yourself might fit into it. ... I still think that the second year is a lot more productive for most``
\end{italicquote}

\subsubsection{Working with external stakeholders}

Many students who had previously not done project work with clients or end-users felt that their project management, client negotiation and communication skills grew at SDA. Some of these interviewees also felt their experience of dealing with clients would have been different were they working as junior developers in the industry where much of the communication and planning is done by project managers or account managers. 

\begin{italicquote}
``I learned a lot about project management. I learned how to evaluate people and tasks and fit those two together. I learned to appreciate the expertise that other people have. I gained a lot of experience in working with people. I learned how to communicate and to conduct myself with clients.`` 
\end{italicquote}

\subsection{Experience compared to capstone courses (RQ 2)}

Most of the interviewees had completed the Software Engineering Capstone Project course during their studies in the CS Bachelor Program (see Table \ref{tab:background-students}). From the student's perspective, the similarity between the capstone and SDA experiences varied greatly. The ones with an external client from another department in the university perceived the work as fairly similar. In their capstone projects, they, too, got work for a real client from the university and developed software for the university. Some even worked with the same technologies and in the same office space during their capstone as they did at SDA. The transition from project work to SDA was fairly small for these students. 

Many, however, considered the difference between the capstone project and SDA as night and day. The considerably longer time they spent working at SDA allowed them to delve fully into the project topics and work on implementing fairly large systems. They also felt a lot more responsible for their work in SDA due to the products being in actual production use by end-users. They referred to capstone projects rarely making it beyond the proof-of-concept phase and rarely getting integrated into the clients' systems. Many of these interviewees made a clear distinction between the other being a school project and the other being real work. They added that there is a certain level of professionalism at SDA that they did not experience while completing the capstone project course.

\begin{italicquote}
``Projects are just projects. I never regarded SDA as a (school) project since SDA is a place of work, and those two are hard to compare. Projects give you experience, but I wouldn't put any school projects into my CV. I can, however, put SDA into the CV since, at SDA, people develop software with a much wider scope, and they do it for actual users and clients.``
\end{italicquote}
\vspace{.4mm}
\begin{italicquote}
``SDA is the only place in the university where you get to work with such projects that match working in the industry``
\end{italicquote}

Out of the interviewees, a group of their own are the six who had been credited with the capstone course due to their participation in SDA. Interestingly, those students generally placed a much higher value on the project course than those who had completed it. Those who had not completed the capstone course speculated it might have been a comparable experience to SDA in terms of realism, client experience and skills it teaches. Those who had completed both the capstone course and SDA regarded SDA as a more valuable experience that could not be entirely replaced by the capstone course.

Regarding the question: \textit{What kind of a course could provide what SDA provides?} 9 out of 15 interviewees stated that they could not see a course could offer the same experience that SDA does. The long duration and the skills and dedication it takes to work at SDA were some of the most severe hindrances to replacing SDA with a course. Students' commitment has a key role in the success of SDA, and the commitment would be hard to replicate in a course environment. The interviewees mentioned that the course would have to last at least a year, including summer, which is hard to achieve without paying the students. Four interviewees hesitated but ended up saying there might be a chance to build something similar, if not quite like it. Those four mainly focused on the skills SDA provides and how they could be taught via project-based courses. 

When asked how many credits they perceive their year at SDA is worth, the interviewees' answers ranged from 15 to 120, averaging around 53 credits. For reference, in the European Credit Transfer System, 1 ECTS credit amounts to approximately 27.5 hours of studying, and a full year is 60 ECTS credits. And even still, many interviewees felt that the salary from SDA was so important that it could not be replaced with course credits.

Another major difference,  
which is hard to replicate, is the employee status and real work experience the students gain when starting to work at SDA. The student interviewees also noted that finding skilled and motivated managers to guide students amidst all their teaching and research responsibilities would become an obstacle to creating such a course.

\begin{italicquote}
``It requires quite a lot of skills from the responsible teacher. The one who coaches and guides the developer team. They ought to be good.``
\end{italicquote}
\vspace{.4mm}
\begin{italicquote}
``Compulsory internship where you go to a real firm. It should be a real work environment to get the same that SDA provides. Even though the software engineering project course tries to mimic it in some ways, people aren't as invested in those projects as they should be in work life. People come and go as they please, and it is different in my experience.``
\end{italicquote}

\subsection{Challenges and further improvement (RQ 3)}

When analysing the interview data, it became clear that some aspects that one interviewee would prefer to change might have been the best thing about SDA to another. High autonomy was the aspect that most %clearly 
divided opinions.  

\begin{italicquote}
%On Q13: \textit{How would you change SDA?} \\
``Of course, if I were given a chance to dictatorially decide how we are gonna do things at SDA, I would start using TDD and pair programming. ... Everything would go through pull requests and code reviews. I mean, \textit{if I got the choice}, I would enforce these strict quality control measures.`` 
\end{italicquote}

At the same other participants replied to question: \textit{What positive aspects do you think SDA has?} by stating:
\begin{italicquote}
``It is free. You get to fly solo. You have some guidelines within which you are free to go solo. 
%You get to choose which technology or library to use for the feature and how to implement it. 
Not a lot of strict practices.``
\end{italicquote}

The most commonly mentioned challenge relates to the experience levels in the group (n=9). SDA is comprised of students with generally little work experience in software development. On the one hand, this was also regarded as highly positive, which meant finding friendships within the work community was easier when the other workers felt more like peers. Having co-workers of the same age group with the same areas of interest also enhanced the feeling of relatedness and created a distinct culture in the workplace. Yet, similar backgrounds cause a certain lack of clear seniority in the group, especially compared to other work environments. In a more typical work environment, there would be senior staff who might have already been in the industry for several years, who then could mentor the junior developers when it comes to work practices, tool selection and even coding patterns. 

\begin{italicquote}
``The good thing about SDA is that you have a lot of responsibility, and you need to do actual things. The thing you don't get that much, which you might get in other jobs, is that there is someone truly experienced and skilled senior developer guiding you. That experience can be educational. So even though you have those ``senior`` coders in the SDA model, it is not the same as someone who has worked in the industry for 15 years. So you are missing that kind of an experience in SDA.``
\end{italicquote}

According to the interviewees, this lack of mentorship sometimes meant that one could continue doing a bad practice for a long time without anyone noticing. Not having assigned mentors and processes in place for mentoring also leaves more responsibility for the progress in the hands of the worker themselves. The interviewees mentioned that one could always easily ask for help from the more senior members of the team. But especially when working remotely, the threshold for asking for help is always higher than the threshold for accepting mentoring when offered by a skilled senior. 

Some suggestions were made on tackling this issue, one being pro-longing the contracts to last more than one year. Another possible solution to this would have more processes or structure within the ways of working, for instance, starting code reviews. Some interviewees mentioned that restricting the autonomy and freedom of a single student might increase the way students learn from each other in the team.

Challenges related to asynchronised working schedules were another repeating theme (n=6). In most cases, this was clearly related to the global COVID-19 pandemic and related remote working arrangements. One interviewee explained how their time before the pandemic was great, but after moving to full remote work, they felt that the peer support and joy of working in a team decreased. Similarly, another mentions that the team's communication and brain-storming were lacking, but a big part of it can be pinned on ``the times we live in``.

Those whose SDA year was entirely done at the office mention teamwork as one of the key learnings from the SDA experience. They also see it as a thing that worked well at SDA. But the ones who worked only remotely did not see teamwork in such a positive light. Or if not negative, it at least was missing from the list of ``on top of your head`` skills that they learned. Although these students felt that the community, friendships and networks they gained through working in SDA were extremely valuable and positive, they still were left with the feeling that something was missing when working only remotely. 

\begin{italicquote}
``The fact that you get to work whenever you want is easier and more flexible, but on the other hand, you don't get the benefits of real teamwork. In my current job, I talk and work with people on a daily basis. But since so many at SDA have the 50\% job contract, the work sometimes gets pretty async. Someone does something, and then you get a message from them, reply when you work and so on. And then we have some weekly meetings. So that is different compared to my current job.``
\end{italicquote}

Finally, whereas the same educational aspects were seen in most interviews, the downsides were much more sparsely reported and SDA was seen mostly as a positive. 

\section{Discussion and Conclusions}
\label{sec:discussion}

We have looked into a novel way of preparing students for the software engineering industry: a university-led internal software startup. This approach was presented via a case study, Software Development Academy (SDA), run at \univ{}. SDA is a software development team comprised of students and led by faculty that develops and maintains educational administrative applications used in the university. A framework for this approach was presented along with the experiences of the students and faculty involved. The startup approach was generally perceived as highly beneficial by the 15 alumni interviewed.

\subsection{Skill gap}

The integral aspects of the SDA (RQ1), also what students saw as valuable in SDA, were discussed in Section~\ref{sec:aspects}. We see an interesting overlap when comparing this with the typical skill gap of software engineering new graduates. For example, in the meta-analysis by Garoussi et al.~\cite{garousi2019closing}, configuration management was the most missed skill in the industry. At the same time, working with software in production, which includes configuration management, was the most widely identified topic in our interviews. The fact that many of the themes emerged after questions like, \emph{what useful skills did you learn while working at the SDA}, hints that the line between the emerged topics and skill gap may be blurred. We believe that at least some of the topics can also be interpreted as the skills what new-graduates often lack (interviewees would have lacked those without SDA).

If the experience from working in a production environment is often challenging to achieve, the obvious question is why this is the case. We hypothesise that, especially in short (i.e., less than a year) industry collaboration periods, it might be intimidating to trust students with the responsibility of a business-critical production environment. In our case, when the staff had already seen the students in previous courses and was able to select the best, we created an environment where students were fully trusted. This was not related to educational software being less important. In SDA, some software had ca. 1000 daily users, and at the time of writing this, it was sold as a service to another university in Finland.

Another knowledge area with high importance and high knowledge gap is \emph{software engineering professional practice}~\cite{garousi2019closing}. This can be compared with our observation that some SDA alumni would have preferred to work with more senior engineers. This is an area where other approaches (e.g., industrial placement~\cite{chillas2015learning}) might pay off with a higher return.

Other connections between the educational aspects of SDA and the typical skill gap were not so explicit, although SDA was experienced as highly industry relevant. In the future, it would be interesting to understand better features of authentic software engineering education (see~\cite{abad2019much}) in general, for which SDA provides one starting point.

\subsection{Comparison to other similar setups}
\label{sec:similars}

Although SDA has some unique characteristics, it can also be compared to other cases when students are hired to work at a university. First, it is common for many research groups to hire students to do software development. However, based on our experiences, research groups typically work on proof-of-concept prototypes that are not meant to be operated and maintained in the long run. This is often the case also in capstone courses with internal clients~\citep{dugan2011survey, buffardi}. Moreover, we argue that the best industrial practices are easy to miss when working as a developer in a research group, and the whole business/product side of the work may be missing since the clients are inside the same group. Next, we describe arrangements that are more similar to SDA.

\citet{ding2017case} present a software engineering capstone project, where students worked on a large university-owned project, a Class Attendance Tracking System (CATS). As the name suggests, the software was designed to help track the class attendance of university students. The project resembles SDA in that it is developed and maintained over a long period of time instead of being a single-shot proof-of-concept by one capstone team. The product owner of CATS is another organisation in the same institution: the staff from the academic enrichment department. The authors state that the product owners were eager to participate in developing the software as it was designed specifically for their needs. Our anecdotal observations from SDA are similar. SDA and CATS are successful examples of student-built software being released to more than ten thousand students. Ding et al.\ also mention that a real and big project such as CATS is very attractive to students. Students in their team found it helpful to put such skills and work in their resumes and found similar job positions in the industry. Similar to our case, Ding et al. thus were also able to witness how using industry-preferred technology in a real university-led project had highly positive effects on the employability of the students.

\citet{williams2021student} report having a very similar software development team as SDA at Berea College. The idea for founding Student Software Development Team (SSDT) rose from the wish to have software matching their needs without running into the prohibitive costs similar to acquiring and maintaining custom-made software from external software houses. They also realised that their version of SDA would provide valuable skill-building experience for students, making them more employable in the software industry. By 2021, SSDT had run for six years and created nine software systems for the purposes of their institution, which marks up to a similar kind of progress done by SDA. In SSDT, the work is organised similarly to SDA in that during summers students work full-time and during the academic terms only part-time. However, in their model, summers are reserved for larger changes and features in the software, and the academic terms are used solely for maintenance, bug fixes and fulfilling small feature requests. In SDA, the nature of work stays the same throughout the year; only the development velocity goes down during the academic term. Similarly to experience gained from SDA, Williams et al.\ also reported the maintenance phase is extremely beneficial for students. They emphasise that maintenance of production-quality software after deployment is a skill rarely taught in software engineering courses. In their experiences, capstone projects and similar are ``completed`` and delivered to customers, with no students' involvement in the maintenance. Their report was written very recently, in 2021. They also mention finding no other institutions hiring students to develop custom software solutions, including maintaining the software throughout its lifetime.

\subsection{Managing internal software startups in universities}

Although this study focused on students' view, some requirements for running an internal startup were also discussed in Section~\ref{sec:case_introduction}. When this is compared with the previous, we  confirm that ``a devoted product owner is critical for the success [of running an internal startup]''~\cite{ding2017case} and 
``consistent communication between the students and the leadership team [is] of the utmost importance.''~\cite{williams2021student}.

Moreover, we have learned that SDA must follow industry trends and best practices to be a credible development organisation in the eyes of both the students and the project clients. This is not an easy task for the management since, when working in a university, the priority is on research and teaching theory-oriented courses. There is simply not enough time or career-related incentives for lecturers and professors to keep in phase with what happens in the industry. In the case of our SDA implementation, the problem is solved as follows. The responsible manager has some years ago taken a leave from the university and worked as a developer in a software startup. Since then, the manager has formed very tight connections to the industry and has several industry partners and mentors helping the manager to keep in touch with the developments in the industry. The manager, together with the industry partners, has also launched a set of open online courses on full-stack software development techniques (see \url{https://fullstackopen.com/en/}) that employ the same software stack and industry-strength operational environment used in SDA. Teaching and maintaining the course set forces the manager to keep in phase with the fast-moving software ecosystem.  The set of courses also ensures that there is plenty of potential students who are known to possess the right skill set for hire. The manager also works with the normal SDA development task from time to time, keeping up the programming routine.

\subsection{How this study changed SDA}

After this study was conducted, we wanted to address the challenges identified. Related to the lack of seniority and scalability, some future students will get an offer for the second year at SDA. Moreover, the steady growth in our project base will likely make it possible to hire one full-time senior(ish) developer to work in SDA. Related to asynchronous schedules, we encourage SDA staff to work at the office and help them plan their work to have more time together. 

\subsection{Threats to validity}\label{validity}

Despite our best effort to be as objective as possible, a close connection with the SDA has likely affected our views. On the other hand, this in-depth understanding of the topic gave us the opportunity to discuss the findings in light of the underlying design principles of the SDA. Moreover, the fact that the interviewers were working at SDA during the time of the interviews, and had also worked as colleagues with some of the interviewees, may have impacted how the interviewees responded. For instance, when asking about the negative aspects of SDA, they may have replied less harshly than if the interviewer was an unknown third party. This bias has been somewhat diminished because when interviews were conducted, none of the interviewees was affiliated with SDA and was thus more likely to express even their negative experiences.

There are many challenges related to implementing SDA elsewhere. First, SDA is led by a faculty member with extensive experience in the software industry. This is likely invaluable in maintaining an authentic software engineering experience, but at the same time, hiring such staff is difficult. Second, the University of Helsinki is a selective research-first university with talented students. However, on top of that, SDA does not rely on exceptional resources and we encourage other faculties to try similar setups. 

\subsection{Conclusions}
\label{sec:conclusions}

When looking at how previous students characterised the Software Development Academy (i.e., RQ1), the following themes emerged: working with production quality software, wide responsibilities and autonomy, employee status and salary, strong community and networking possibilities, and easy integration with other studies. Among these, the possibility to work in a production environment and maintenance of widely used software is an opportunity often missed in software engineering education.

Compared to most capstone courses (i.e., RQ2), SDA lasts longer, and is more intensive. Instead of prototypes, students create production-quality code and also operate and maintain applications that have been created by previous SDA student generations. The scalability of the SDA concept depends heavily on how much time and effort the faculty members are willing to use to ensure that SDA runs smoothly. As scaling the concept might be a challenge, we should seek how similar features (i.e., topics identified when answering RQ1) could be integrated into software education (e.g., capstone courses). Moreover, it would also be interesting to see if this framework could be extended to other disciplines.

Finally, when looking at the downsides of the approach (RQ3), two themes emerged: lack of clear seniority and asynchronised working schedules. The latter can amplify other challenges or even downplay some of the benefits (e.g., a strong community). Thus, to get the best out of the limited time of seniors, internal software startups should address remote working cautiously.  

In the future, we would like to examine also the perspective of other entities involved, such as the client organisations within the university and employers of those students who enrolled in SDA to measure whether the skills they acquired gave them any edge or advantage over other college graduates.

\appendix

\section{Interview questions -- student perspective}
\label{appendix:student_interview}

\begin{description}
\item[\textbf{Theme 1.}]{\textbf{Interviewee background}}
\end{description}

\begin{itemize}
\item How did you end up working in SDA?

\item How did your studies proceed before SDA? %How did your studies proceed before your time in the SDA?
\item How did your studies proceed during SDA?
\item How many years of software engineering experience did you have before starting at SDA?
\item Can we connect your answers to the data of your course completions?

\end{itemize}

\begin{description}
\item[\textbf{Theme 2.}]{\textbf{Experiences from Software Development Academy}}
\end{description}

\begin{itemize}
\item What positive aspects SDA has?
\item What about negative aspects?
\item What kinds of tasks did you have while working in SDA?
\item How many credits do you think your time at SDA was worth? ie. What would be an approriate amount of credits to cover SDA experience?
\item How would you change SDA?
\item How did the nature of work change during your year in SDA?

\end{itemize}

\begin{description}
\item[\textbf{Theme 3.}] {\textbf{Comparing SDA with project courses}}
\end{description}

\begin{itemize}
\item How was your experience of SDA in comparison to your experience of project-based courses?
\item What positive and negative aspects do you think SDA has, compared to project-based courses?
\item How much responsibility do you think working in SDA has, compared to working in project-based courses?
\item What kind of a project course could offer the same as SDA?
\item What other differences do you feel that SDA has, compared to project-based courses?
\end{itemize}

\begin{description}
\item[\textbf{Theme 4.}]{\textbf{Learning useful skills}}
\end{description}

\begin{itemize}
\item What useful skills did you learn while working at SDA?
\item What skills did you learn during project courses, which have been proven to be useful in your work-life?
\item What skills did you learn during project courses, which have been proven to be useful in your work-life?
\item Which skills were such, that you could not have gotten them from elsewhere in the university?
\end{itemize}

\begin{description}
\item[\textbf{Theme 5.}]{\textbf{Work-life experience}}
\end{description}

\begin{itemize}
\item How did SDA affect your employment?
\item How does your current work-life differ from your work-life in SDA?
\item How did the traditional project-based courses affect your employment?
\item What other useful courses can you think of, which have taught you skills beneficial in your work-life? what skills did you learn through them?

\end{itemize}

\balance
\bibliographystyle{IEEEtranN}
% Loading bibliography database

\bibliography{main}

\end{document}